\documentclass[journal=jacsat,manuscript=article]{achemso}
\usepackage{achemso}
\usepackage{multirow}
\usepackage[version=3]{mhchem}
\usepackage[english]{babel}
\usepackage{filecontents}
\usepackage{graphicx}
\usepackage{courier}
\usepackage{amssymb,amsmath,amsthm,mathrsfs,comment,afterpage,accents}
\usepackage{mathtools}
\usepackage{braket}
\usepackage{mathrsfs}
\usepackage{courier}
\makeatletter
\def\@dotsep{4.5}
\makeatother
\usepackage{dcolumn}
\usepackage{bm}
\usepackage{hyperref}
\usepackage[capitalize]{cleveref}
\renewcommand\vec\mathbf

\setkeys{acs}{maxauthors = 0}
\setkeys{acs}{articletitle = true}
\SectionNumbersOn

\title{Excited States via Coupled Cluster Theory without Equation-of-Motion Methods:
Seeking Higher Roots with
Application to Doubly Excited States and Double Core Hole States
}
\author{Joonho Lee}
\email{linusjoonho@gmail.com}
\author{David W. Small}
\author{Martin Head-Gordon}
\email{mhg@cchem.berkeley.edu}
\affiliation{
Department of Chemistry, University of California, Berkeley, California 94720, USA
Chemical Sciences Division, Lawrence Berkeley National Laboratory, Berkeley, California 94720, USA
}
\begin{document}
\maketitle
\newpage

\begin{abstract}
In this work, we revisited the idea of using the coupled-cluster 
ground state formalism 
to target excited states.
Our main focus was
targeting 
doubly excited states and double core hole states.
Typical equation-of-motion (EOM) approaches for obtaining these states struggle 
without higher-order excitations than doubles.
We showed that by using a non-aufbau determinant optimized via the maximum overlap method 
the CC ground state solver can target higher energy states.
Furthermore, just with singles and doubles (i.e., CCSD), we demonstrated that
the accuracy of $\Delta$CCSD and $\Delta$CCSD(T) far surpasses that of EOM-CCSD for doubly excited states.
The accuracy of $\Delta$CCSD(T) is nearly exact for doubly excited states considered in this work.
For double core hole states, 
we used an improved ansatz for greater numerical stability by freezing core hole orbitals. 
The improved methods, core valence separation (CVS)-$\Delta$CCSD and CVS-$\Delta$CCSD(T), were
applied to the calculation of the double ionization potential of small molecules.
Even without relativistic corrections, we observed qualitatively accurate results with CVS-$\Delta$CCSD and CVS-$\Delta$CCSD(T).
Remaining challenges in $\Delta$CC include the description of open-shell singlet excited states with the single-reference CC ground state formalism 
as well as excited states with genuine multi-reference character. 
The tools and intuition developed in this work may serve as a stepping stone 
towards directly targeting arbitrary excited states using ground state CC methods.
\end{abstract}
\newpage
\section{Introduction}
A conceptually simple approach to solving the Schr{\"o}dinger equation is
to diagonalize the Hamiltonian represented by the many-particle basis set spanning the entire Hilbert space.
While this full configuration interaction (FCI) approach (or exact diagonalization) is formally exact,
it becomes quickly unfeasible due to the exponentially growing dimension of the Hilbert space\cite{szalay2011multiconfiguration}.

Coupled-cluster (CC) theory, which is usually limited to singles and doubles (i.e., CCSD), has been a popular approximate solver to the Schr{\"o}dinger equation.
Unlike truncated CI methods, truncated CC methods are size-consistent and therefore can be reliably applied to large systems and reach the thermodynamic limit.
Most of the CC applications have been focused on approximating the ground state of systems and therefore CC methods are usually considered to be ground state methods \cite{bartlett2007coupled}.

There is a way to compute excitation energies of CC wavefunctions based on the equation-of-motion (EOM-CC)\cite{krylov2008equation} formalism or the linear response (LR-CC)\cite{koch1990excitation} formalism. The most widely used method is EOM-CCSD which provides very accurate single-excitation gaps. The accuracy of EOM-CCSD for valence single excitations is about 0.1-0.2 eV. However, EOM-CCSD commonly fails to predict double-excitation gaps, and the typical error is about 1 eV or even greater than this. 
These failures could be avoided if the desired excited state is in a different irreducible representation from that of the ground state since one could just employ
a ground state CCSD calculation. However, if there is no point group symmetry in the system or the desired state is in the same irreducible representation, this
workaround is no longer an option.
The failure of EOM-CCSD for doubly excited states is largely due to the lack of relaxation of doubles amplitudes which can be usually achieved by having triple excitations (i.e., EOM-CCSDT)\cite{watts1994inclusion,kucharski2001coupled}. Since EOM-CCSDT has a cost which scales $\mathcal O(N^8)$, much research has been dedicated to improving the double-excitation gaps of EOM-CCSD by approximating the effect of connected triples either via an $\mathcal O(N^8)$ scaling method with a smaller prefactor or
an $\mathcal O(N^7)$ scaling method. These methods include
EOM-CCSDT-n \cite{watts1995economical,watts1996iterative}, EOM-CCSD(T) \cite{watts1995economical}, EOM-CCSD($\tilde{\text{T}}$) \cite{watts1996iterative}, EOM-CCSD(T') \cite{watts1996iterative}, CC3 \cite{koch1995excitation,christiansen1995response}, and CCSDR(3)\cite{christiansen1996perturbative}.

Another challenging class of excited states for EOM-CC are core-ionized states.
In most cases, these core ionization energies can be well described by the EOM with ionization potential (EOM-IP) approaches\cite{stanton1994analytic,krylov2008equation}.
However, one has to obtain a large number of eigenvectors to cover the energy range for core ionizations which can be very time-consuming for an $\mathcal O(N^5)$ method. There are tricks to remedy this problem to an extent via 
core valence separation (CVS)\cite{cederbaum1980many,barth1985theoretical,wenzel2014calculating,wenzel2015analysis,vidal2019new}, but it does not solve the inherent drawbacks of EOM-IP-CCSD.
In other words, CVS-EOM-IP-CCSD fails when EOM-IP-CCSD fails. 
In particular, for some core-ionized states, EOM with excitations up to doubles is not sufficient.

Recently, there have been increasing interest in so-called $\Delta$SCF methods \cite{triguero1998calculations,ehlert2018efficient,evangelista2013orthogonality,derricotte2015simulation,gilbert2008self,barca2018simple} as an alternative to the linear-response mean-field approaches such as CI singles (CIS) and time-dependent density functional theory (TDDFT)\cite{Dreuw2005}. 
In this category, the most popular approach is based on the maximum overlap method (MOM) developed by Gilbert and Gill \cite{gilbert2008self}. 
The resulting approximate excited states from $\Delta$SCF are not orthogonal to the approximate ground state.
This seems to be suboptimal since the exact excited state should be orthogonal to the exact ground state. 
However, extensive benchmarks have so far suggested that the non-orthogonality of approximate wavefunction methods are not problematic to get good energetics. Furthermore, it is possible to diagonalize the Hamiltonian with those non-orthogonal determinants to obtain orthogonal states in the end. 
This approach is called non-orthogonal CI (NOCI) \cite{thom2009hartree, sundstrom2014non}. 

Similar in spirit to $\Delta$SCF, it is possible to obtain approximate solutions to exact excited states using the CC wavefunction parametrization. We call this approach $\Delta$CC, and this is the focus of our work.
In $\Delta$CC, one computes the ground state CCSD (gd-CCSD) and an excited state CCSD (ex-CCSD) energies and takes
a difference between them to compute the corresponding excitation gap.
Performing an ordinary ground state CCSD calculation on an excited reference determinant leads to a desired ex-CCSD energy.
Just like $\Delta$SCF targets an excited SCF solution, $\Delta$CC targets an excited CC solution that starts from an excited reference state.
We emphasize that $\Delta$CC is not a new approach and has been known in literature for a while\cite{
Meissner1993,
Jankowski1994,
Jankowski1994a,
Jankowski1995,kowalski1998towards,
Jankowski1999,
Jankowski1999a,
Jankowski1999b,
piecuch2000search,
Podeszwa2002,
Podeszwa2003,
mayhall2010multiple}.
In particular, there are seminal works by Kowalski and co-workers that attempt to find higher roots in CC methods using the homotopy method \cite{kowalski1998towards,piecuch2000search}. 
They also established connections 
between these roots and excited states in FCI for model systems such as \ce{H4}. 
$\Delta$CC has been underappreciated because of the obscure nature of CC amplitude solutions.
In particular, the higher roots of the CC amplitude equation are difficult to assign to a specific state.
It is also often very difficult to converge the CC amplitude equation and multiple CC roots sometimes correspond to the same FCI state\cite{mayhall2010multiple}.

While these drawbacks make $\Delta$CC not so appealing in general, we will show that $\Delta$CC can be an accurate tool for
excited states that are dominated by one Hartree-Fock (HF) state. CCSD with perturbative triples (CCSD(T)) is a {\it de facto} standard method for
the ground state of systems with one dominant determinant. One may expect CCSD(T) to work well as long as the underlying electronic structure has only one dominant determinant, which does not need to be the ground state.
In such cases we expect the excited state CCSD(T) energies to be quite accurate and even similar in quality to that of the ground state calculation.
We found excited states dominated by one double-excitation to be a perfect candidate for this approach.
This is largely because the state assignment becomes much easier since it is dominated by one determinant.
As mentioned earlier, EOM-CCSD fails to describe such states with dominant double-excitations so $\Delta$CC can be an excellent alternative
with the same $\mathcal O(N^6)$ cost.

It is also worthwhile to note that $\Delta$CC has been used in the literature to compute core ionization energies \cite{Nooijen1995,Ohtsuka2006,Besley2012,Zheng2019}.
Similarly to the double excitations, this is due to the ease of assigning proper states as well as relatively more stable amplitude iterations.
The amplitude convergence can often become problematic, but this issue can be completely removed by a CVS-like treatment which freezes core hole orbitals as proposed in ref. \citenum{Zheng2019}.
The resulting CVS-$\Delta$CC is a good computational tool for targeting core-ionized states at the cost of ground state CCSD calculations while retaining the full flexibility of the CC wave function.
In this work, we will focus on the computation of double ionization potentials which currently not many methods are able to compute. In particular, the CVS implementation of EOM-DIP-CCSD \cite{Wladyslawski2002,sattelmeyer2003use} is unavailable at the time of writing this manuscript.
Furthermore, we will illustrate that EOM-DIP-CCSD does not retain the full flexibility of CCSD and it is not an exact approach for computing electronic energies for 2-electron systems when starting from a 4-electron reference.

The goal of this paper is to (1) revive the idea of $\Delta$CC with the emphasis on targeting doubly excited states and double core hole states and (2) present numerical data on small molecules to support this idea.

\section{Theory}
\subsection{Coupled-Cluster Theory as an Arbitrary Root Solver}
Coupled-cluster (CC) wavefunctions use an exponential parametrization,
\begin{equation}
|\Psi\rangle = e^{\hat{T}} |\Phi_0\rangle
\end{equation}
where $\hat{T}$ is the CC cluster operator defined as
\begin{equation}
\hat{T} = \sum_\mu t_\mu \hat{\tau}_\mu
\end{equation}
with $\hat{\tau}_\mu$ being the excitation operator which creates $|\Phi_\mu\rangle$ from the reference determinant $|\Phi_0\rangle$
and $t_\mu$ is the cluster amplitude.
The CC ansatz then follows
\begin{equation}
\hat{\mathcal{H}} |\Psi\rangle = E_\text{CC} |\Psi\rangle
\end{equation}
assuming that $|\Psi\rangle$ is an eigenstate of $\hat{\mathcal H}$ and $E_\text{CC}$ is the corresponding eigenvalue (i.e., energy),
\begin{equation}
E_\text{CC} = \langle \Phi_0 | \hat{H} | \Psi \rangle
\end{equation}
The amplitudes $\{t_\mu\}$ are obtained by solving
\begin{equation}
t_\mu E_\text{CC} = \langle \Psi_\mu | \hat{\mathcal{H}} |\Psi\rangle
\label{eq:amp}
\end{equation}
Up to this point, we have not assumed whether we are trying to approximate the ground state or one of the excited states.
In fact, the only assumption that has been made is that the state $|\Psi\rangle$ is an eigenstate of a given Hamiltonian.

The bias towards the reference state $|\Phi_0\rangle$ built in the exponential parametrization
controls which state we are targeting.
The exponential parametrization is expanded to
\begin{equation}
e^{\hat{T}} = \hat{1} + \hat{T} + \frac{\hat{T}^2}{2!} + \cdot\cdot\cdot.
\end{equation}
Since non-strongly correlated systems typically have amplitudes smaller than 1,
the largest component in a usual CC wavefunction is the reference state $|\Phi_0\rangle$.
This is known for model problems due to the work by Kowalski and co-workers \cite{kowalski1998towards}.
However, with the advances in $\Delta$SCF methods \cite{gilbert2008self},
it is meaningful to revisit this idea for more complex chemical systems.
We will denote such excited CC states as ex-CC states where as the ground state CC state will be referred to as gd-CC.
The energy difference between gd-CC and ex-CC states defines
the $\Delta$CC approach 
for electronic excitation energies.
This viewpoint can also be easily extended to number-changing excitations such as ionization potential (IP) and
electron attachment (EA).


It is important to note two main limitations of these $\Delta$CC methods for electronic excited states (EE), IP, and EA. 
First, traditional CC (TCC) methods are not capable of describing 
strong electron correlation so $\Delta$TCC methods are limited to 
states of single-reference character.
Those with multi-reference character need 
more sophisticated CC approaches 
that can handle strong correlation.
Examples of such approaches include CC valence bond with singles and doubles (CCVB-SD) \cite{Small2012,Lee2017},
parametrized CCSD (pCCSD) \cite{Huntington2010}, distinguishable cluster SD (DCSD) \cite{Kats2013}, etc.
For the purpose of this paper, we will focus on 
the application of TCC approaches
to states described well by a single determinant.
Applying more advanced CC approaches to multi-reference problems will be an interesting
topic for future study. We will refer to TCC simply as CC for the rest of this paper.

Second, the computation of transition properties such as oscillator strengths and transition dipole moments 
is not straightforward and seems to scale exponentially with system size.
Any transition properties between gd-CCSD and ex-CCSD states
should technically involve a CC state for both bra and ket (first-order derivatives for each).
Moreover, orbitals of gd-CCSD are not orthogonal to any orbitals of ex-CCSD in general.
The evaluation of transition properties therefore formally scales exponentially with system size
if done exactly.
This contrasts with EOM approaches where the bra state is not a CC state, instead it is only a linear wavefunction
with the same set of orbitals as the gd-CCSD state.
One may consider linearizing both of the CC states to evaluate transition properties to get an approximate answer,
but the exact evaluation of such properties is still highly desirable.
For the purpose of this work, we will compute only energies and leave the computation of transition properties to future
study.

\subsection{Equation-of-Motion Coupled-Cluster Theory}
For a given ground state CC wavefunction,
one can solve a Hamiltonian eigenvalue problem in the linear response space.
We first define the CC Lagrangian,
\begin{equation}
\mathcal L (\mathbf \lambda, \mathbf t) = 
\langle
\tilde{\Psi}(\lambda)|\hat{H}
|\Psi (\mathbf t)
\rangle_C
\label{eq:lag}
\end{equation}
where the subscript $C$ implies that it involves only ``connected'' diagrams \cite{Shavitt2009}
and
the bra is defined as
\begin{equation}
\langle
\tilde{\Psi}(\lambda)|=
\langle
\Phi_0|
(1 + \hat{\Lambda}) 
\end{equation}
with
the deexcitation operator $\hat{\Lambda}$ being
\begin{equation}
\hat{\Lambda} = \sum_\mu \lambda_\mu \hat{\tau}_\mu^\dagger,
\end{equation}
%
Evidently, we have $\mathcal L = E_\text{CC}$ for $\mathbf t$ such that the CC amplitude equation (\cref{eq:amp}) is satisfied.
Then, the equation-of-motion (EOM) Hamiltonian (or the CC Jacobian) can be derived from the linear response of this Lagrangian\cite{koch1990excitation},
\begin{equation}
J_{\mu\nu} = 
\frac{\partial
 \mathcal L (\mathbf \lambda, \mathbf t)
 }
 {\partial \lambda_\mu \partial t_\mu}
 \bigg|_{\mathbf t=\mathbf t_0}
\label{eq:eom}
\end{equation}
where $\mathbf t_0$ is a set of amplitudes that satisfies the ``ground state'' CC amplitude equation.
Since $\mathcal L$ is a linear function of $\lambda$, $\mathbf J$ is independent from $\lambda$.
EOM is linear response because it is a derivative of an energy expression with respect to the wavefunction parameter for both bra and ket.

In EOM-CCSD, \cref{eq:eom} is formed in the space of singles and doubles. 
Evidently, EOM-CCSD cannot describe any excited states that mainly contain triples and higher excitations.
What may not be immediately obvious is that EOM-CCSD, in practice, cannot describe excited states with
strong double excitation character.
In our view, there are two aspects of \cref{eq:eom} that should be highlighted:
(1) orbitals are determined for the ground-state SCF calculation and are fixed,
and (2)
the CC amplitudes, $\mathbf t$, are also determined for  the ground state and are also fixed.
This naturally imposes constraints on EOM calculations and
reducing the effects of those constraints requires higher-order excitations (in this case triples).
This has, of course, been well-known in the community and a method such as EOM-CC(2,3) is motivated by this observation\cite{Hirata2000}.
EOM-CC(2,3) takes the ground state CCSD wavefunction and forms the CC Jacobian in the space of singles, doubles, and triples. 
This is not really a linear-response method since it goes beyond the ground state parameter space, but it has shown
to improve the accuracy of EOM-CCSD greatly especially for states with strong double excitation character.
As we will see later, without any non-perturbative connected triples, $\Delta$CCSD and $\Delta$CCSD(T) can
perform significantly better than EOM-CCSD.

This formalism can be extended to Fock space to treat numbers of electrons different from that of the ground state. 
The method that is relevant to the present work is the EOM ionization potential (EOM-IP) methods. 
In EOM-IP-CCSD, the ``singles'' operator (1p) removes an electron and the ``doubles'' operator (1h2p) removes two electrons from occupied orbitals and adds an electron to one of the unoccupied orbitals. By performing EOM-IP-CCSD on an $N$-electron system, one can obtain the energies of the corresponding $(N-1)$-electron system and therefore ionization potentials. 
Removing an electron from a molecule must be accompanied by sufficient orbital relaxation. This is implicitly done by the 1h2p operator which resembles the singles operator for $(N-1)$ systems.
Interestingly, EOM-IP-CCSD effectively has only ``singles''-type excitations from a $(N-1)$ reference state via the 1h2p operator and no higher excitations.
Therefore, the flexibility of EOM-IP-CCSD is smaller than that of CCSD or EOM-CCSD in terms of describing correlation between electrons.

A similar conclusion can be drawn for EOM double IP (EOM-DIP) methods.
EOM-DIP-CCSD employs the ``singles'' operator (2p) which removes two electrons and the ``doubles'' operator (1h3p) which removes three electrons from occupieds and adds an electron back to virtuals.
From an $(N-2)$ reference state, this EOM-CC state effectively has only ``singles'' excitations.
A majority of those singles would account for orbital relaxation and only little correlation effect would be gained from using EOM-DIP-CCSD. 

The limited flexibility of EOM-DIP-CCSD can be most clearly understood by considering a model problem that contains four electrons and four orbitals.
If we apply EOM-DIP-CCSD to this system, one would generate some determinants within the two-electron Hilbert space but not all.
\begin{figure}[h!]
\includegraphics[scale=0.5]{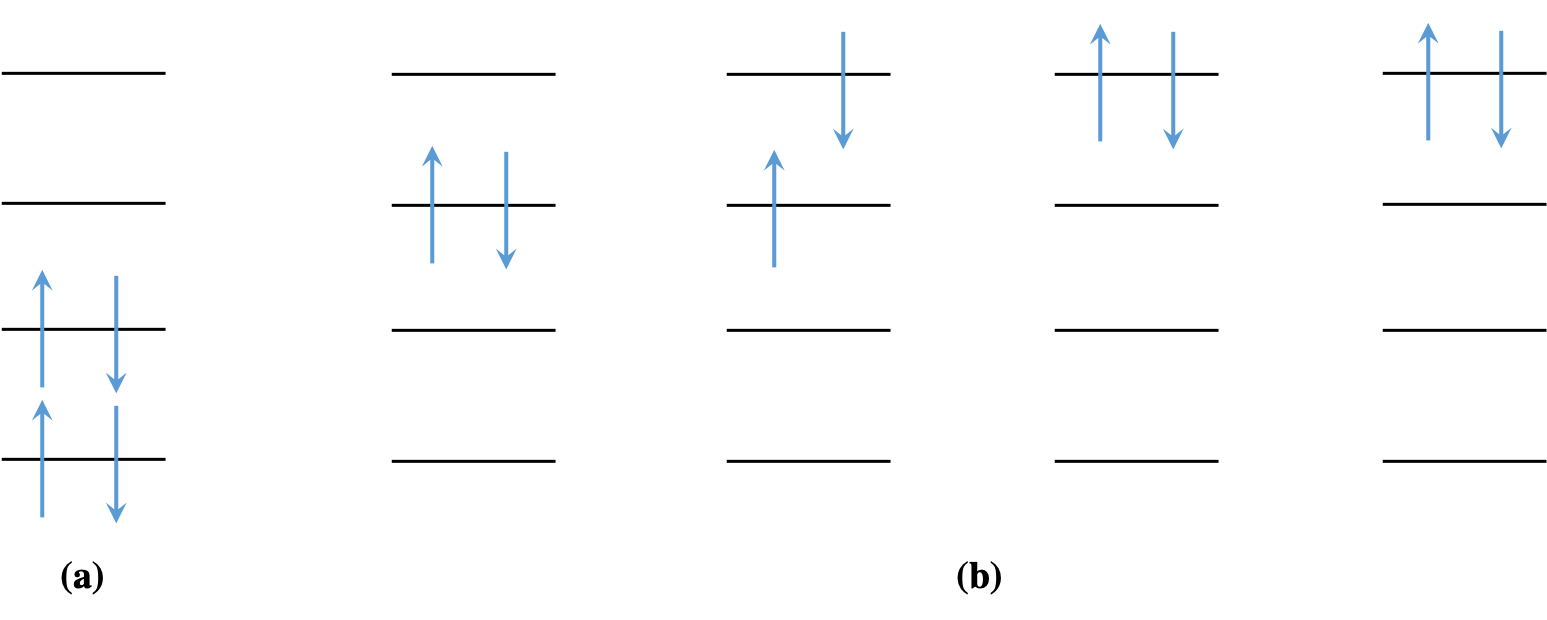}
\caption{
\label{fig:eomdip}
(a) Reference determinant for a 4-electron system and
(b)
four determinants in the 2-electron sector
that are unreachable via EOM-DIP-CCSD.
They are each 2p4h excitations from the reference, but EOM-DIP-CCSD allows at most 1p3h excitations.
}
\end{figure}
In \cref{fig:eomdip} (b), we explicitly show four determinants that are unreachable via EOM-DIP-CCSD if one uses the ground state determinant with four electrons shown in \cref{fig:eomdip} (a).
This is somewhat disappointing because CCSD is exact for 2-electron systems. EOM-DIP-CCSD is not exact for 2-electron systems when starting from a 4-electron reference.
On the other hand, if one were to compute DIPs of a 4-electron system via $\Delta$CCSD, at least the 2-electron system energy is exactly treated via CCSD. The remaining error is then solely from the CCSD error in the ground state.



\section{\ce{H2}: a proof-of-concept example }
For the ground state of \ce{H2}, CCSD is exact since it includes
all possible excitations of two electrons in the system.
Similarly, EOM-CCSD is exact for every state of \ce{H2}
and therefore with conventional ground-state CCSD (i.e., gd-CCSD) and EOM-CCSD 
one can get all of the electronic states of \ce{H2} exactly for a given basis set.
We will show that it is possible to reproduce those exact energies with the excited-state CCSD (i.e., ex-CCSD) method
without severe numerical issues.

\begin{figure}[h!]
\includegraphics[scale=0.75]{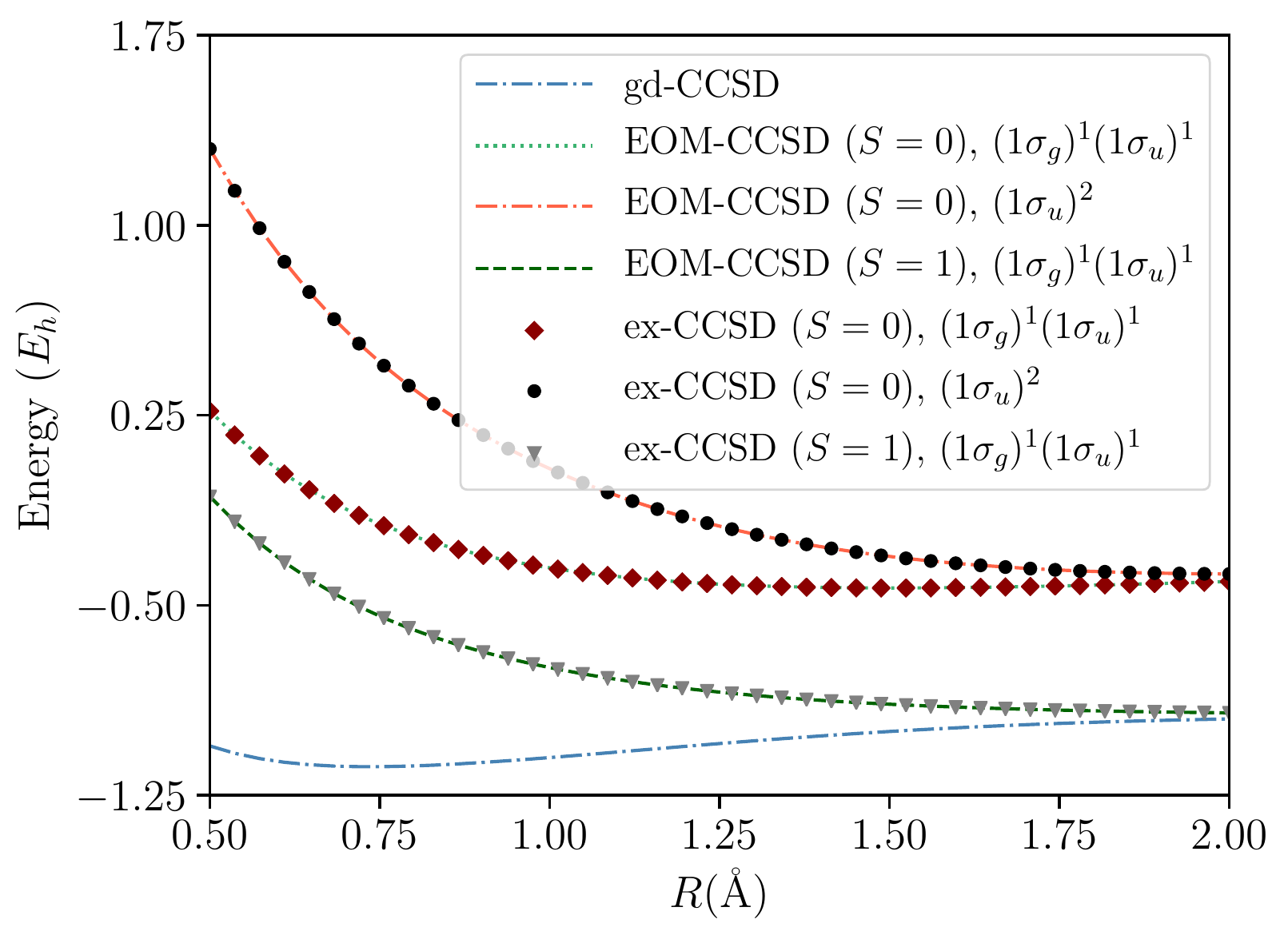}
\caption{
\label{fig:h2sto3g}
All of the states with $M_S=0$ of \ce{H2} in the STO-3G basis set computed with gd-CCSD, EOM-CCSD, and ex-CCSD.
There are two $S=0$ excited states (labeled by 1 and 2) and one $S=1$ state.
Note that ex-CCSD follows EOM-CCSD exactly for all of the excited states.
}
\end{figure}
\begin{figure}[h!]
\includegraphics[scale=0.65]{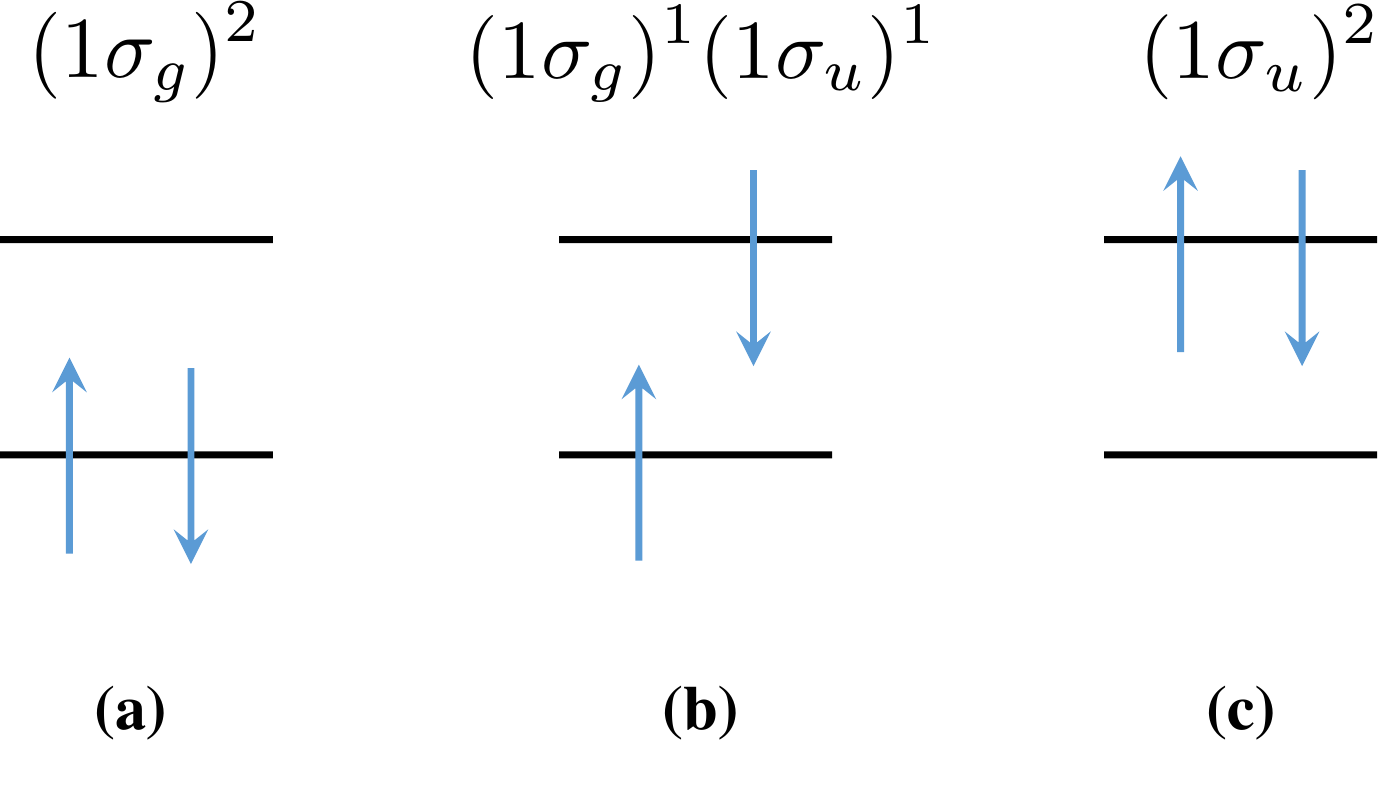}
\caption{
\label{fig:h2guess}
Reference determinants to obtain the (a) gd-CCSD, (b) ex-CCSD ($S=0$),1 and ex-CCSD ($S=1$), and (c) ex-CCSD ($S=0$), 2
energies in \cref{fig:h2sto3g}.
}
\end{figure}

First, in \cref{fig:h2sto3g}, we present the results for \ce{H2} with the STO-3G basis set.
With this basis set, there are only four states in the $M_S=0$ sector.
All of these states can be obtained by running CCSD calculations with a carefully chosen reference determinant along with initial guess amplitudes. For the ground state and the doubly excited state $(1\sigma_u)^2$, the MP2 amplitude guess was used.
For the singly excited states (singlet and triplet), we use a guess of $T_{hh}^{ll} = \pm1$, respectively, where $T_{hh}^{ll}$ denotes the doubles amplitude for $1\sigma_g \rightarrow 1\sigma_u$.
This strategy was enough to obtain the numerical data presented in \cref{fig:h2sto3g}.

\begin{figure}[h!]
\includegraphics[scale=0.75]{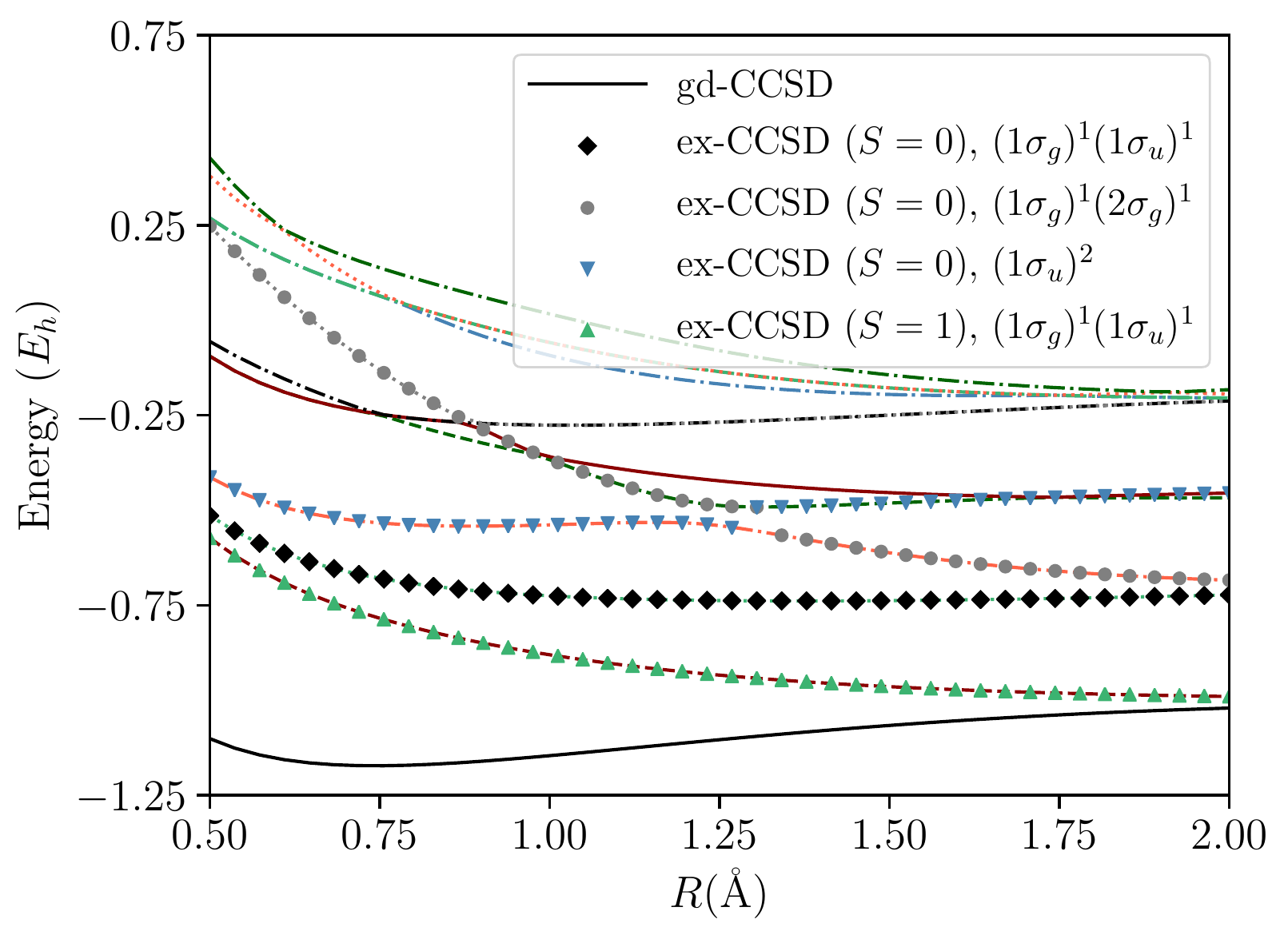}
\caption{
\label{fig:ccpvtz}
Electronic energies of \ce{H2} with cc-pVTZ:
the singlet ground state from gd-CCSD, three singlet excited states from ex-CCSD, ten EOM-CCSD singlet states, one triplet excited state from ex-CCSD, and
the triplet ground state from ex-CCSD and EOM-CCSD for $M_S=0$.
Note that EOM-CCSD states are plotted with lines and they are not labeled for simplicity.
ex-CCSD follows a root in EOM-CCSD in all cases.
}
\end{figure}

The same principles can be applied to a larger basis set calculation (cc-pVTZ)\cite{Dunning1989} as shown in \cref{fig:ccpvtz}. 
A distinct feature of the targeted ex-CCSD method is
that it follows a root of the same character throughout potential energy surface (PES).
This is most obvious from the ex-CCSD state obtained from the $(1\sigma_g)^1(2\sigma_g)^1$ reference shown in \cref{fig:ccpvtz}.
At $R=0.5$\:\AA\:, it starts as the 6-th excited state of EOM-CCSD, stays on the same state, and
eventually becomes the third excited state of EOM-CCSD as the bond gets elongated.
Around $R=1.25$\:\AA\: an avoided crossing appears between the third and fourth excited states.
The ex-CCSD energies for these two states switch near this avoided crossing.
This is natural for a targeted excited state since it follows a state of desired character.

\section{Applications to Doubly Excited States}\label{sec:des}
\subsection{\ce{CH2} $(1{}^1A_1\rightarrow 2{}^1A_1)$}
Methylene (or carbene) has a triplet ground state.
The singlet ground state ($1{}^1A_1$) for \ce{CH2} is therefore an excited state.
We optimized the geometry of \ce{CH2} on this electronic surface with $\omega$B97X-D and the def2-QZVPPD basis set.
Interestingly, the next excited state with the same term symbol (i.e., $2{}^1A_1$) has strong double excitation character.
This doubly-excited state is dominated by a closed-shell single determinant and therefore it is a 
perfect candidate for the $\Delta$CC methods.
Furthermore, it is possible to perform brute-force methods such as the semi-stochastic heat-bath CI (SHCI) method and a second-order perturbation correction (SHCI+PT2) on this system \cite{Smith2017}. As such,
we compare $\Delta$CCSD and EOM-CCSD against near-exact SHCI results.
We employed the frozen-core approximation for the results presented in this section.

\begin{figure}[h!]
\includegraphics[scale=0.45]{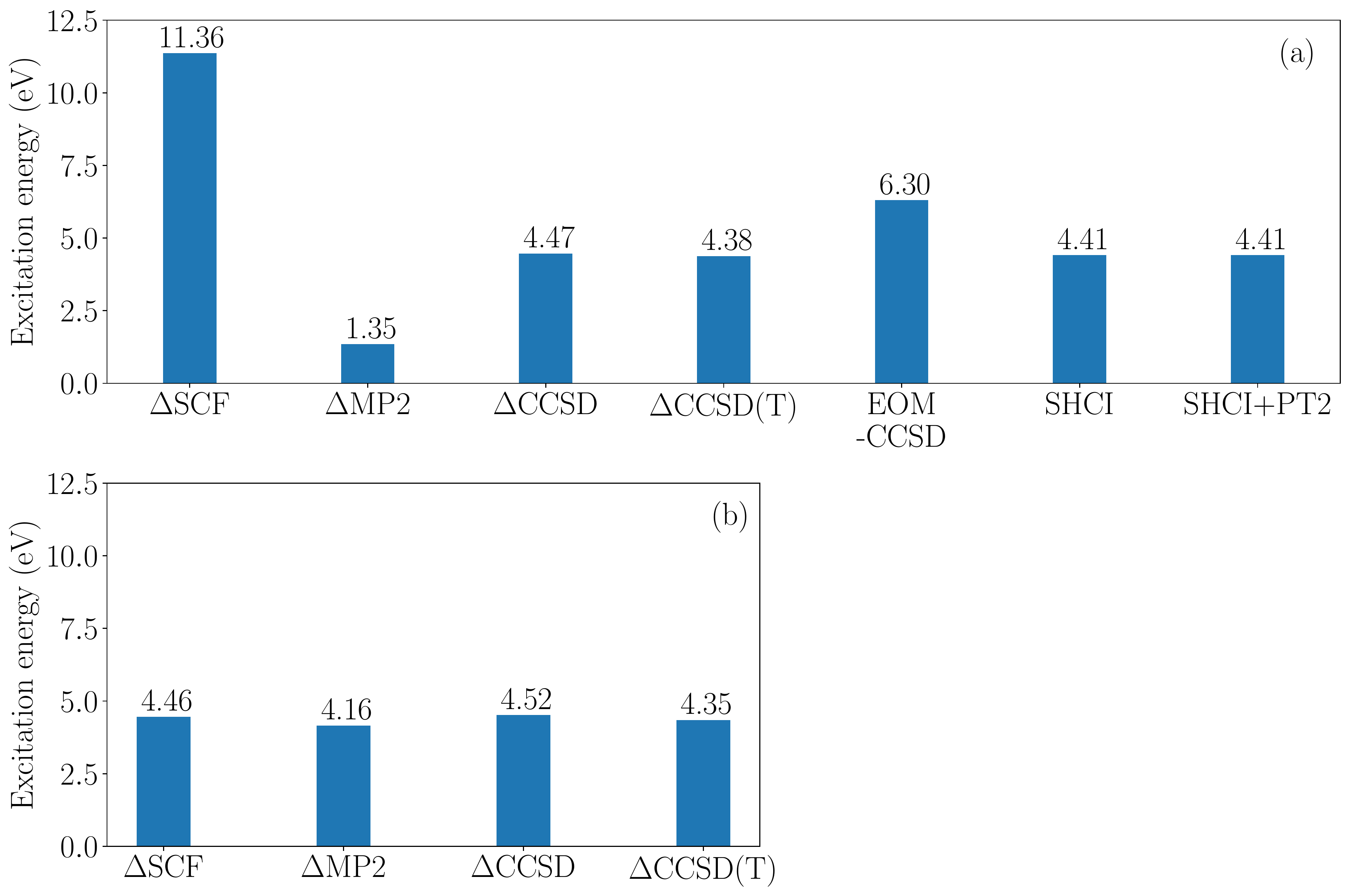}
\caption{
\label{fig:ch2}
The $(1{}^1A_1\rightarrow 2{}^1A_1)$ excitation energies of \ce{CH2}
obtained from various methods with the aug-cc-pVQZ basis set. The excited states for the $\Delta$SCF, $\Delta$MP2, and $\Delta$CCSD
methods
are based on a non-Aufbau state using (a) ground state orbitals and (b) a metastable SCF state optimized via MOM, respectively.
The statistical error associated with SHCI+PT2 is 0.00032 eV which is negligible on the scale of these plots.
}
\end{figure}

In \cref{fig:ch2}, the excitation energies for the $(1{}^1A_1\rightarrow 2{}^1A_1)$ transition are presented for various methods computed with the aug-cc-pVQZ basis set.
For $\Delta$ methods, we used a reference that doubly-occupies the $1{}^1A_1$ LUMO.
In other words, we used
a reference with a transition of $(3a_1)^2$$\rightarrow$$(1b_2)^2$.
As shown in \cref{fig:ch2} (a), the use of this non-Aufbau state with ground state orbitals yields ill-behaved $\Delta$SCF and $\Delta$MP2 energies.
This erratic behavior does not appear in the case of $\Delta$CCSD and $\Delta$CCSD(T) due to the singles operator. 
EOM-CCSD shows an error of 1.89 eV which is much larger than its typical error for valence single excitations.
In contrast, $\Delta$CCSD and $\Delta$CCSD(T) show remarkably accurate excitation energies whose errors are less than 0.1 eV.
This is because the $2{}^1A_1$ state is mainly dominated by one closed-shell determinant which can be 
accurately described by CCSD.

In \cref{fig:ch2} (b), we examine the effect of orbital relaxation for the $\Delta$ methods.
The non-aufbau determinant was subsequently optimized to lower the HF energy using the MOM algorithm\cite{gilbert2008self}.
$\Delta$SCF and $\Delta$MP2 improve significantly when using an orbital-optimized excited state determinant.
However, in the case of $\Delta$CCSD and $\Delta$CCSD(T), the results are more or less the same as before.
This is expected because the important effect of orbital optimization can be incorporated through single excitations.

\subsection{Loos and co-workers' benchmark set: ethylene and formaldehyde}
With advances in brute-force approaches,
it is now possible to produce high-quality benchmarks for small molecules.
An example of such benchmarks is Loos and co-workers' recent study
where
they used 
a brute-force selected CI (sCI) approach \cite{Loos2019} to produce reference energies for 
doubly excited states of a total of 14 small molecules:
acrolein, benzene, beryllium, carbon dimer, carbon trimer, ethylene, formaldehyde, glyoxal, hexatriene, 
nitrosomethane, nitroxyl, pyrazine, and tetrazine.

Interestingly, most of these molecules exhibit multi-reference character in the doubly excited state
and they are challenging for single-reference CCSD to describe properly.
For instance, Loos and co-workers considered the $2s^2 \rightarrow 2p^2$ excitation in Be.
This state has three equally important determinants, $2p_x^2$, $2p_y^2$, and $2p_z^2$.
Therefore, this state is of multi-reference by nature and not easy to describe with CCSD.
More complications arise for butadiene and its isoelectronic species, acrolein and glyoxal.
These molecules have a low-lying dark state that
has three significant configurations, one of which is a doubly-excited configuration.
This dark state is likewise beyond the scope of conventional single-reference CCSD.

On the other hand, the doubly excited states of ethylene and formaldehyde 
were found to be well described by a single determinant.
Therefore, these are perfect candidates for the $\Delta$CC approach.
Given the \ce{CH2} results discussed above, we obtained the excitation gaps for $\Delta$ methods with optimized ground-state and
excited-state orbitals.

In the following benchmarks, we shall compare our results against benchmark numbers reported by Loos and co-workers\cite{Loos2019}.
For smaller basis sets, they produced near-exact excitation gaps based on
sCI with second-order correction (sCI+PT2) and extrapolated full CI (exFCI) methods\cite{Loos2018}.
This should be adequate in assessing the quality of $\Delta$ methods for small basis sets.
For larger basis sets, Loos and co-workers produced EOM-CC3 excitation energies.
As we will see, EOM-CC3 is less accurate than $\Delta$CCSD(T) when compared to sCI+PT2 and exFCI in smaller basis sets.
Nevertheless, EOM-CC3 is a widely used iterative $\mathcal O(N^7)$ correlated excited state method that can yield qualitatively correct excitation gaps for doubly excited states.
As such, we will also compare $\Delta$CC methods to EOM-CC3.

\begin{figure}[h!]
\includegraphics[scale=0.2325]{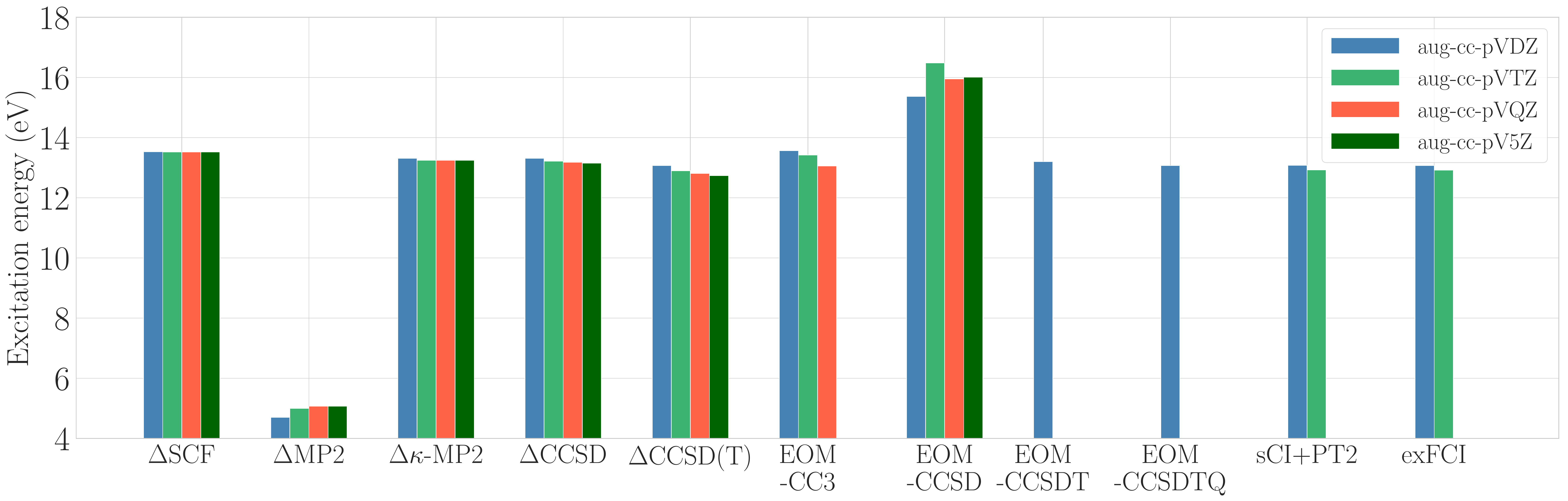}
\caption{
\label{fig:eth}
The $(1{}^1A_{g}\rightarrow 2{}^1A_g)$ excitation energies of ethylene for various basis sets.
EOM-CC3, EOM-CCSDT, EOM-CCSDTQ, sCI+PT2 and exFCI results were taken from ref. \citenum{Loos2019}.
The error bars on ex-FCI are 0.01 eV and 0.06 eV for aug-cc-pVDZ and aug-cc-pVTZ basis sets, respectively.
}
\end{figure}

In the case of ethylene, as shown in \cref{fig:eth}, 
EOM-CCSD significantly overestimates the gap by 2-3 eV. 
This highlights the failure of EOM-CCSD for doubly excited states. 
The doubles amplitudes, $\mathbf R_2$, in EOM-CCSD are not enough to describe this state and
it is necessary to incorporate triple excitations to reach reasonable accuracy as for instance in EOM-CCSDT with aug-cc-pVDZ.
The role of quadruples is relatively unimportant in this case.
The use of a reference determinant, $(1b_{1u})^2\rightarrow (1b_{2g})^2$, yields
remarkably accurate excitation energies with $\Delta$CCSD and $\Delta$CCSD(T).
$\Delta$CCSD(T) excitation energies are within the error bar of exFCI. Compared to sCI+PT2, the errors are 0.01 eV and 0.03 eV for aug-cc-pVDZ and aug-cc-pVTZ basis sets, respectively.
This is better in accuracy than EOM-CC3 whose error is 0.49 eV for both basis sets.

Since the doubly excited state of ethene is well described by a single determinant, relatively accurate gaps from $\Delta$SCF 
are not unexpected. What is surprising is the striking underestimation of the gap in $\Delta$MP2.
One would think that for problems for which a single determinant is qualitatively correct MP2 should perform well.
While this is commonly true, in the case of $\Delta$MP2, the orbital optimization of a non-aufbau determinant often leads to a very small gap between highest occupied molecular orbital (HOMO) and lowest unoccupied molecular orbital (LUMO)\cite{Yost2018}. Consequently, the MP2 correlation energy for such determinants would be heavily overestimated (i.e., more negative than it should be). 
As an attempt to remedy this problem, we applied a recently developed regularized MP2 method ($\kappa$-MP2)\cite{lee2018regularized,lee2019distinguishing,lee2019two,bertels2019third}.
Since small energy gaps will be damped away, the resulting correlation energy is stable even for those non-aufbau determinants. 
As shown in \cref{fig:eth}, $\Delta$$\kappa$-MP2 excitation energies are similar to those of $\Delta$CCSD, 
which highlights the utility of $\kappa$-MP2 for excited state simulations.

\begin{figure}[h!]
\includegraphics[scale=0.2325]{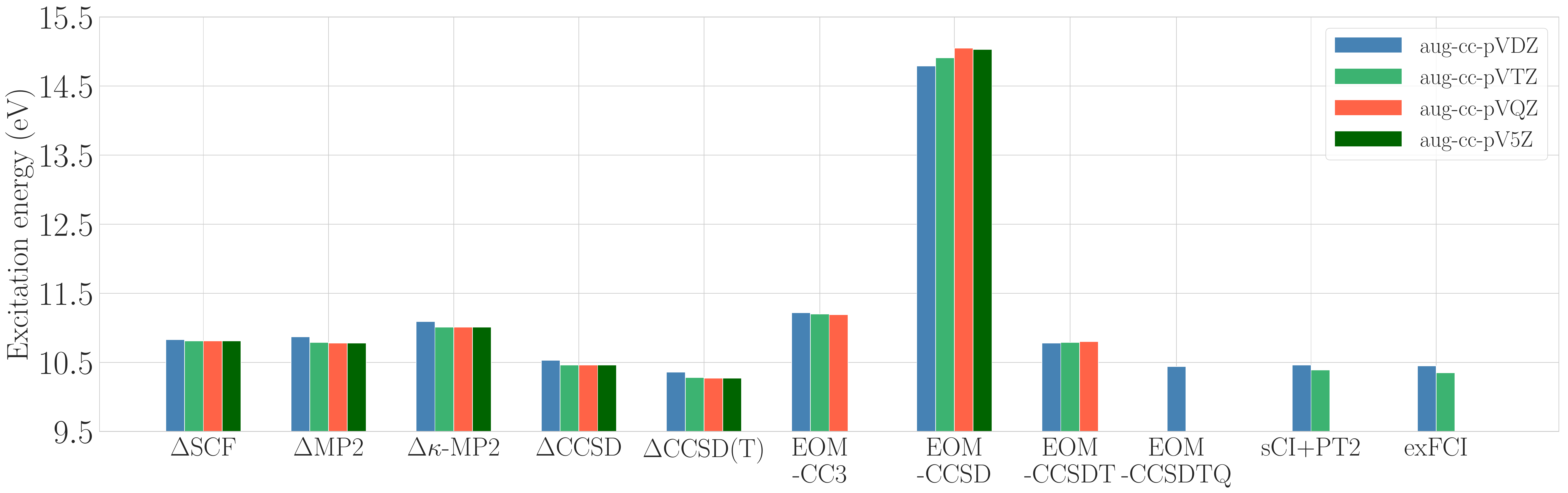}
\caption{
\label{fig:form}
The $(1{}^1A_1\rightarrow 3{}^1A_1)$ excitation energies of formaldehyde for various basis sets.
EOM-CC3, EOM-CCSDT, EOM-CCSDTQ, sCI+PT2 and exFCI results were taken from ref. \citenum{Loos2019}.
The error bars on ex-FCI are 0.01 eV and 0.03 eV for aug-cc-pVDZ and aug-cc-pVTZ basis sets, respectively.
}
\end{figure}

In \cref{fig:form}, we present another successful application of $\Delta$CC methods.
The doubly excited state of formaldehyde is largely dominated by one determinant.
Similarly to previous examples, EOM-CCSD significantly overestimates the excitation gap. 
The error of EOM-CCSD is about 4-5 eV in this case.
EOM-CCSDT greatly improves but incorporating quadruples (i.e., EOM-CCSDTQ) is necessary to reach near-exact results.
Directly targeting the excited state with CCSD and CCSD(T) using a non-aufbau reference determinant, ($(2b_1)^2\rightarrow(2b_2)^2$), handles this state nearly exactly.
With aug-cc-pVTZ, $\Delta$CCSD and $\Delta$CCSD(T) yield an error of 0.07 eV compared to sCI+PT2.
$\Delta$CCSD overestimates whereas $\Delta$CCSD(T) underestimates the gap.
This is better than EOM-CC3, which overestimates the gap by 0.81 eV.
Given the accuracy of $\Delta$CCSD(T), we conclude that the role of connected quadruples in describing
this state can be made negligible with a properly chosen reference deterimnant. 
We also note that $\Delta$SCF produces a qualitatively correct gap and $\Delta$MP2 does not 
exhibit the overcorrelation problem previously shown in the case of ethylene. 
Therefore, $\Delta$$\kappa$-MP2 does not offer any improvement. In fact, $\Delta$$\kappa$-MP2 performs about 0.2 eV worse than $\Delta$MP2.
\subsection{Summary}
In summary, not every doubly excited state requires an explicit treatment for triples unlike what was stated in Loos and co-workers' work \cite{Loos2019}. 
It is only those states that are dominated by more than one determinant, which require a more sophisticated treatment than single-reference CCSD and CCSD(T).
For doubly excited states with one dominant determinant, we showed that CCSD and CCSD(T) can directly target such states by simply employing a non-aufbau determinant as a reference state.
The errors of $\Delta$CCSD and $\Delta$CCSD(T) were found to be less than 0.1 eV for the systems considered in this work.

\section{Applications to Double Core Hole States}
Core-ionized states are another class of excited state 
that can be effectively handled by $\Delta$CC methods.
In fact, this was noted in the literature several times \cite{Nooijen1995,Ohtsuka2006,Besley2012,Zheng2019} and
was recently revived by Zheng and Cheng \cite{Zheng2019}.
In particular, Zheng and Cheng benchmarked single core hole (SCH) states for various small molecules and found about 0.13 eV standard deviation for $\Delta$CCSD(T) in the ionization energies with respect to experimental values.
Interested readers are referred to ref. \citenum{Zheng2019} for further information about their work.

What we will focus in this work is the use of $\Delta$CCSD(T) for double core hole (DCH) states.
Following the prescription by Zheng and Cheng for SCH states, we first obtain an $(N-2)$ electron reference state and
freeze two unoccupied core orbitals for numerical stability.
The removal of unoccupied core orbitals is similar in spirit to the CVS\cite{cederbaum1980many,barth1985theoretical,wenzel2014calculating,wenzel2015analysis,vidal2019new} treatment in EOM-CC
and it explicitly prevents the CC wavefunction from collapsing to the ground state of the same number of particles. In our case, a double excitation from HOMO to the unoccupied core orbitals would yield much lower energy than the desired core-ionized state.
This is the source of numerical instability.
The approach which freezes core hole orbitals will be referred to as CVS-$\Delta$CC.

Investigating DCH states to probe chemical environment was first proposed by Cederbaum and co-workers\cite{Cederbaum1986,Cederbaum1987,Ohrendorf1991}. 
Compared to SCH states, DCH states are much more sensitive to chemical environment. A classic example that illustrates this point is
the series of hydrocarbons, \ce{C2H2}, \ce{C2H4}, and \ce{C2H6}\cite{Cederbaum1986}. 
Creating a SCH state by removing an electron from a carbon atom in these molecules 
results in IPs that differ only by tens of eV from each other.
On the other hand, DIPs exhibit a difference over 4 eV or so per C-C bond. 
This highlights the utility of DCH states in probing chemical environment.
Since Cederbaum's proposal, DCH states have also been experimentally realized \cite{
Cryan2010,
Fang2010,
Linusson2011,
Lablanquie2011,
Lablanquie2011a,
Berrah2011,
Salen2012,
Penent2015,
Goldsztejn2016}. In particular, two-site DCH (TSDCH) states are sensitive to chemical structure so obtaining TSDCH states in experiments have become a focus\cite{Salen2012,Lablanquie2011a}. A single-site DCH (SSDCH) state can be readily obtained from a closed-shell $(N-2)$ reference determinant whereas 
TSDCH states are inherently of open-shell singlet character. 
In this section we will study both kinds of DCH states and apply the $\Delta$CC approach to obtain their electronic energies. 

The algorithm to obtain DCH states in CVS-$\Delta$CC is as follows:
\begin{enumerate}
\item 
Perform an SCF calculation on an $N$-electron system.
\item
Localize core orbitals (and optionally valence orbitals separately) if there is more than one atom for the chemical element of interest.
We employed Boys localization\cite{Boys1960} for this step and other localization schemes are also possible.
\item Identify core orbitals that will be made to be unoccupied.
\item Remove two electrons from hand-selected core orbitals and perform
an $(N-2)$-electron SCF calculation using the MOM algorithm or Newton's method.
\item Perform a CCSD calculation on the converged $(N-2)$-electron reference.
Note that it is necessary to freeze the two core hole orbitals in this step for numerical stability.
\end{enumerate}
We note that the CVS approach naturally requires frozen-core calculations for the $N$-electron ground state calculation.

\subsection{Single-site double core hole states}
We will investigate the SSDCH states of five small molecules, \ce{CO}, \ce{CH4}, \ce{NH3},  \ce{N2}, and \ce{CO2} and compare
the DIP values computed from CVS-$\Delta$CCSD and CVS-$\Delta$CCSD(T) with those of experiments. 
All geometries were obtained from geometry optimization with $\omega$B97X-D\cite{Chai2008} and aug-cc-pCVTZ\cite{Dunning1989,Woon1995}.

\begin{table}[h]
\begin{tabular}{c|c|c|c|c|c|c}\hline
Molecule & Ionization & Basis Set & $\Delta$SCF & CVS-$\Delta$CCSD & CVS-$\Delta$CCSD(T) & Exp. \\ \hline
\multirow{3}{*}{\ce{CO}} & \multirow{3}{*}{C 1s$^{-2}$} & aCVTZ & 667.55 & 665.88 & 665.76 & \multirow{3}{*}{668(4)} \\  
 &  & aCVQZ & 667.24 & 665.36 & 665.20 &  \\
  &  & aCV5Z &  667.20 &   665.29 & 665.12  &  \\ \hline
  \multirow{3}{*}{\ce{CH4}} & \multirow{3}{*}{C 1s$^{-2}$} & 
  aCVTZ &  650.77 & 650.59  & 650.64   & \multirow{3}{*}{651.5(5)} \\  
  &  & aCVQZ & 650.49  & 649.88  & 649.92  &  \\
  &  & aCV5Z & 650.45  &  649.74 &  649.77 &  \\ \hline
  \multirow{3}{*}{\ce{NH3}} & \multirow{3}{*}{N 1s$^{-2}$} & 
  aCVTZ &   890.86 &  891.22  &   891.33  & \multirow{3}{*}{892.0(5)} \\  
  &  & aCVQZ &  890.54  & 890.53   &  890.77  &  \\
  &  & aCV5Z & 890.49  & 890.37  & 890.48  &  \\ \hline
    \multirow{3}{*}{\ce{N2}} & \multirow{3}{*}{N 1s$^{-2}$} & 
  aCVTZ &   901.18  & 901.75    & 901.83     & \multirow{3}{*}{902.6(5)} \\  
  &  & aCVQZ &  900.84   &    901.18 &  901.24   &  \\
  &  & aCV5Z &  900.79 &  901.06 &   901.13&  \\ \hline
  \multirow{3}{*}{\ce{CO}} & \multirow{3}{*}{O 1s$^{-2}$} & aCVTZ & 1174.74 & 1175.92 & 1176.23 & \multirow{3}{*}{1178.0(8)} \\
 &  & aCVQZ & 1174.32 & 1175.23 & 1175.54 &  \\
  &  & aCV5Z & 1174.25 & 1175.08 & 1175.39 &  \\\hline
      \multirow{3}{*}{\ce{CO2}} & \multirow{3}{*}{O 1s$^{-2}$} & 
  aCVTZ &     1172.16 & 1172.48     &   1172.62    & \multirow{3}{*}{1173(2)} \\  
  &  & aCVQZ &   1171.75   &    1171.81  &   1171.96   &  \\
  &  & aCV5Z &   1171.68 &  1171.67  &  1171.81  &  \\ \hline
\end{tabular}
\caption {
Double ionization potentials (eV) for single-site double core hole states.
aCVXZ (X=T, Q, 5) is a short form for aug-cc-pCVXZ. Experimental values are obtained from refs. \citenum{Lablanquie2011,Salen2012}.
}
\label{tab:SSDCH}
\end{table}

In \cref{tab:SSDCH}, we present DIPs for SSDCH states using $\Delta$CC methods with increasing the size of basis set.
In the case of ionizing two electrons from a carbon atom, we observe roughly 2 eV of correlation effects in \ce{CO}.
On the other hand, the correction effect plays a smaller role for \ce{CH4}.
In both molecules, increasing the size of the basis set reduces the DIPs. 
\ce{CO} is well within the error bar of the experimental value
partly because the experimental error bar is quite large. For \ce{CH4}, CVS-$\Delta$CCSD and CVS-$\Delta$CCSD(T) exhibit an error on the order of 1 eV. Due to the lack of relativistic effect treatment in our calculations, this error is not so surprising \cite{Tashiro2010} and we will leave more thorough benchmarks for future study. Nonetheless, $\Delta$SCF, CVS-$\Delta$CCSD, and CVS-$\Delta$CCSD(T)
all yield the correct trend that CO's DIP is several eVs higher than \ce{CH4}'s DIP. This qualitative conclusion holds even at the SCF level.

For SSDHC states involving nitrogen core vacancies, we investigated \ce{NH3} and \ce{N2}. Their experimental estimates are about
10 eV apart. Similar to the previous cases, we observe smaller DIPs with larger basis sets. With the aug-cc-pCV5Z basis set, we observe about 1.5 eV error for all the methods. The result improves as we go from SCF to CVS-$\Delta$CCSD(T).
A major source of error is again the lack of relativistic effects.
Nevertheless, all these methods successfully capture qualitative differences between these two chemical species. Namely, the DIPs of \ce{NH3} and \ce{N2} differ by about 10 eV.

In the case of oxygen, we investigated \ce{CO} and \ce{CO2}. 
The DIPs of these molecules
were experimentally shown to 
differ by about 5 eV. This qualitative difference is well described by all of the methods examined here.
However, the quantitative agreement between CVS-$\Delta$CC methods and experimental values 
was only within several eVs as before in other systems.

In passing, we note that we neglected strong correlation present in double core hole states.
Namely, in \ce{N2}, there are two possible ways to obtain a DCH state on a nitrogen atom.
Likewise, there are two equally important choices for \ce{CO2} for generating a DCH state on an oxygen atom.
One may think that this would require mixing of two such references. In fact, this effect was studied in the context of NOCI with singles for simulating core-valence excitations in our group \cite{oosterbaan2018non}. Given the quantitative agreement between CVS-CCSD(T) and experimental values for IPs reported in ref. \citenum{Zheng2019}, we suspect that such strong correlation effect may not be important in simulating core-ionized states. This can also be found in experimental results where single core hole states are usually localized on one atom even when there are multiple atoms of the same chemical element \cite{Jolly1984}.

\subsection{Two-site double core hole states}
Unlike SSDCH, TSDCH states are all open-shell states.
Two core holes are created at different atomic sites and therefore 
two unpaired open-shell electrons will remain. There are two 
possible ways to spin-couple these two electrons: singlet and triplet.
We will obtain rough energetics of these states by employing a broken-symmetry HF reference state 
whose $\langle S^2 \rangle$ is close to 1.0.

Such a reference state is well-suited for Yamaguchi's approximate spin projection (AP)\cite{Yamaguchi1988}. With AP, 
one can obtain a spin-pure energy for the singlet state. Namely, 
\begin{equation}
E_{S=0}  = \frac{E_\text{BS} - (1- \alpha) E_{S=1}}{\alpha}
\label{eq:Eproj}
\end{equation}
where $E_\text{BS}$ is the broken-symmetry state energy, $E_{S=0}$ is the singlet energy,
$E_{S=1}$ is the triplet energy, and the coefficient $\alpha$ is given by
\begin{equation}
\alpha = \frac{\langle S^2\rangle_{S=1} - \langle S^2 \rangle _\text{BS}}{\langle S^2\rangle_{S=1} - \langle S^2\rangle_{S=0}}.
\end{equation}
which uses the $\langle S^2\rangle$ values of BS and $S=1$ states (assuming $\langle S^2\rangle_{S=0}=0$).
Clearly, \cref{eq:Eproj} requires not only a broken-symmetry calculation but also a $M_S=1$ calculation to obtain $E_{S=1}$ and $\langle S^2\rangle_{S=1}$.
Within our CVS approach, an $M_S=1$ TSDCH calculation would require a different number of frozen core and virtual orbitals for $\alpha$ and $\beta$ spin sectors. This is uncommon to run for most quantum chemistry packages available at the moment and therefore we will leave the use of AP for TSDCH states for future study.

The TSDCH states of four small molecules, \ce{CO}, \ce{CO2}, \ce{N2}, and \ce{N2O}, are investigated.
We report the DIP values computed from CVS-$\Delta$CCSD and CVS-$\Delta$CCSD(T) and compare them with those of experiments in \cref{tab:TSDCH}. 
All geometries were obtained from geometry optimization with $\omega$B97X-D and aug-cc-pCVTZ.

\begin{table}[h]
\begin{tabular}{c|c|c|c|c|c|c}\hline
Molecule & Ionization & Basis Set & $\Delta$SCF & \shortstack{CVS-\\$\Delta$CCSD} & \shortstack{CVS-\\$\Delta$CCSD(T)} & Exp. \\ \hline
\multirow{3}{*}{\ce{CO}} & \multirow{3}{*}{C 1s$^{-1}$, O 1s$^{-1}$} & 
aCVTZ & 853.77 (1.31) & 854.94 (1.15) & 854.96 & \multirow{3}{*}{855(1)} \\
 &  & aCVQZ & 853.58 (1.32) & 854.75 (1.15) & 854.76 &  \\
  &  & aCV5Z & 853.54 (1.32) & 854.72 (1.15) & 854.73 &  \\\hline
 \multirow{3}{*}{\ce{CO2}} & \multirow{3}{*}{C 1s$^{-1}$, O 1s$^{-1}$} & 
aCVTZ & 851.71 (1.15) & 851.68 (1.19) & 851.51 & \multirow{3}{*}{849(1)} \\
 &  & aCVQZ & 851.53 (1.15) & 851.46 (1.19) & 851.28 &  \\
  &  & aCV5Z & 851.49 (1.15) & 851.42  & 851.24 &  \\\hline
  \multirow{3}{*}{\ce{N2}} & \multirow{3}{*}{N 1s$^{-1}$, N 1s$^{-1}$} & 
aCVTZ & 834.06 (1.16) & 835.34 (0.38) & 835.50 & \multirow{3}{*}{836(2)} \\
 &  & aCVQZ & 833.90 (1.16) & 835.12 (0.39) & 835.27 &  \\
  &  & aCV5Z & 833.86 (1.16) & 835.07 (0.39) & 835.22 &  \\\hline
   \multirow{3}{*}{\ce{N2O}} & \multirow{3}{*}{N 1s$^{-1}$, N 1s$^{-1}$} & 
aCVTZ & 836.27 (1.25) & 834.99 (1.23) & 834.49 & \multirow{3}{*}{834(2)} \\
 &  & aCVQZ & 836.11 (1.26) & 834.82 (1.23) & 834.30 &  \\
  &  & aCV5Z & 836.07 (1.26) & 834.79  & 834.27 &  \\\hline
\end{tabular}
\caption {
Double ionization potentials (eV) for two-site double core hole states.
For $\Delta$ methods, the numbers in parentheses indicate the corresponding $\langle S^2\rangle$ value.
aCVXZ (X=T, Q) is a short form for aug-cc-pCVXZ.
Experimental values were taken from ref. \citenum{Salen2012}.
}
\label{tab:TSDCH}
\end{table}

First, we study the TSDCH states where one core hole is localized on carbon and the other one is localized on oxygen in \ce{CO} and \ce{CO2}. Experimentally, these two molecules have DIPs that are 6 eV apart from each other. 
The difference is quite small at the SCF level as two values are only 2 eV apart.
With CVS-$\Delta$CCSD, the difference becomes 3.2 eV and CVS-$\Delta$CCSD(T) yields a difference of 3.5 eV.
While these are not quantitatively accurate, they all still correctly reproduce the qualitative behavior observed experimentally.

Next, we investigate the TSDCH states in \ce{N2} and \ce{N2O} by creating one core hole on each nitrogen.
The DIPs of these two molecules are only 2 eV apart and almost within the experimental error bar from each other. 
Nonetheless, our goal is to reproduce the fact that the DIP of \ce{N2} is slightly larger than the DIP of \ce{N2O}.
At the SCF level, the trend is reversed. With $\Delta$SCF, \ce{N2} has a DIP that is 2 eV lower than that of \ce{N2O}.
With CVS-$\Delta$CCSD and CVS-$\Delta$CCSD(T), 
a correct trend is reproduced. At the CCSD level, the DIP of \ce{N2} is only 0.3 eV higher than that of \ce{N2O} whereas
the difference becomes 0.9 eV at the CCSD(T) level.

As we can see, even without the relativistic treatment and spin-projection, we observe good qualitative agreement between CVS-$\Delta$CC methods and experiments. It will be valuable to revisit these systems with proper relativistic corrections and Yamaguchi's AP and try to observe a quantitative agreement between theory and experiments. We note that $\langle S^2\rangle$ at both SCF and CCSD levels lies between 0 and 2 which asserts that these states are suitable for Yamaguchi's AP.

\subsection{Summary}
In this section, we applied the $\Delta$CC method to both single-site and two-site double core hole states. 
Similarly to the doubly excited states studied in \cref{sec:des}, there was no difficulty encountered as long as the underlying CC state we are targeting is single-reference by nature. 
Furthermore, the satisfactory numerical stability was ensured by freezing core holes for CC calculations. 
The resulting CVS-$\Delta$CC methods were tested on a variety of small molecular systems.
While it was difficult to make a quantitative comparison between CVS-$\Delta$CC and experiments due to the lack of relativistic treatment and large error bars in experimental values, both CVS-$\Delta$CCSD and CVS-$\Delta$CCSD(T) captured qualitative trends observed in experiments even when $\Delta$SCF failed to do so. 
Furthermore, given small differences between CVS-$\Delta$CCSD and CVS-$\Delta$CCSD(T), it appears that the role
of electron correlation can be fully captured at the CVS-CCSD level.
We were not able to make comparisons to other available approaches because EOM-DIP-CCSD cannot obtain
those highly excited core hole states at a reasonable cost.
Furthermore, a production-level CVS-EOM-DIP-CCSD implementation is currently unavailable. 

\section{Conclusions}
In this work, 
we revisited the long-standing idea 
of using coupled-cluster (CC) wavefunction
to directly target excited states
that may be beyond the scope of equation-of-motion (EOM) approaches.
In particular, we focused on using CC with singles and doubles (CCSD) 
to describe (1) doubly excited states
and (2) double core hole states.

For doubly excited states, we show that it is possible to directly target an excited state through the ground state formalism of CCSD
without numerical difficulties as long as the targeted state is dominated by one determinant.
We achieve this simply by employing a non-aufbau reference determinant that is orbital-optimized at the mean-field level
via the maximum overlap method.
A directly targeted CCSD and CCSD with perturbative triples (CCSD(T)) excited state was shown to yield excellent excitation gaps
for \ce{CH2}, ethylene, and formaldehyde. In particular, $\Delta$CCSD(T) was shown to yield near-exact excitation gaps when compared to brute-force methods in a small basis set. 
This is quite promising since EOM-CCSD typically exhibits an error greater than 1 eV for these states.
Furthermore, $\Delta$CCSD(T) was found to be more accurate than EOM third-order approximate CC (EOM-CC3).

Likewise, double core hole states (DCHs) can be directly obtained from the ground state CCSD formalism.
This is also done by using a non-aufbau reference determinant that has double core holes.
To ensure numerical stability, those core holes were frozen in correlation calculations.
The resulting $\Delta$CC ansatz is referred to as core-valence separation (CVS)-$\Delta$CC and 
was benchmarked over double ionization potentials (DIPs) of small molecular systems (\ce{CO}, \ce{CO2}, \ce{N2}, \ce{N2O}, and \ce{NH3}).
Without relativistic corrections, CVS-$\Delta$CCSD and CVS-$\Delta$CCSD(T) were not able to reach quantitative accuracy when compared to experimental values. Nonetheless, they were able to estimate correct trends even when the mean-field method ($\Delta$SCF) could not.

With the success of $\Delta$CC described here, some interesting new directions become apparent.
A more thorough investigation of open-shell singlet states in conjunction with Yamaguchi's spin-projection will be interesting.
Currently, for valence excitations, we investigated states dominated by a closed-shell determinant.
Two-site DCHs were investigated without spin-projection. 
With spin-projection, a broader class of states will be accessible and spin-pure energies for two-site DCHs can be obtained.
Secondly, the use of more sophisticated CC methods such as CC valence bond with singles and doubles (CCVB-SD) 
for targeting excited states with multi-reference character will be interesting. 
Lastly, in addition to core-ionized states, targeting core-valence excited states will be a promising candidate to apply 
the techniques described in this work.
Some of these are currently underway in our group.

\section{Acknowledgements}
This work was supported by the Director, Office of Science, Office of Basic Energy Sciences of the U.S. Department of Energy under contract no. DE-ACO2-05CH11231.
We thank Anna Krylov for stimulating discussions on useful applications of $\Delta$CC for core-ionized states.
J. L. thanks Soojin Lee for constant encouragement.
\newpage
\bibliography{deltacc}
\bibliographystyle{achemso}
\end{document}